# Egocentric vision IT technologies for Alzheimer disease assessment and studies

Hugo Boujut[1], Vincent Buso[1], Guillaume Bourmaud[2], Jenny Benois-Pineau[1], Rémi Mégret[2],
Jean-Philippe-Domenger[1], Yann Gaestel[3], Jean-François Dartigues[3]

[1] LaBRI, UMR 5800 CNRS/University of Bordeaux 1, Bordeaux, France
[2] IMS, UMR 5218 CNRS/University of Bordeaux 1, Bordeaux, France
[3] ISPED, U897 INSERM, Bordeaux, France

*Abstract* − *Egocentric vision technology consists in capturing the actions of persons from their own visual point of view using wearable camera sensors. We apply this new paradigm to instrumental activities monitoring with the objective of providing new tools for the clinical evaluation of the impact of the disease on persons with dementia. In this paper, we introduce the current state of the development of this technology and focus on two technology modules: automatic location estimation and visual saliency estimation for content interpretation.*

## I. INTRODUCTION

The context of actual work is the multi-disciplinary research on Alzheimer disease [1, 3, 4]. The goal here is to ensure an objective assessment of the capacity of patients in conducting IADLs (Instrumental Activities of Daily Living). Impairments in IADL are considered as the core of the diagnosis and the management of Alzheimer's disease. Thus an objective evaluation is crucial.

Within the Dem@care FP7 EU[1] project several complementary approaches are combined. This project follows our previous study on the applicability of the technology and automatic recognition of activities ANR Blan IMMED project [1, 3] Based on a combination of several sensors (within a subset of: motion, presence, fixed or wearable cameras, audio…) automatic processing of the acquired data extracts low-level events, that are then analyzed and compared to provide a comprehensive representation of the activities of the monitored person. This information can then be shown to a clinician or medical staff as indices to better assess the status of a patient or fed into an automatic system for patient feedback.

In this paper, we consider the use of a wearable camera worn on the shoulder as a tool that provides a privileged viewpoint on the IADLs. The complete analysis by a clinician of the videos acquired in such situations would help assessing the effect of the disease by providing a direct close-up observation of the IADLs when the patient accomplish them, either in an ecological situation at home or during directed exercises in the Lab. This nevertheless requires the development of advanced automatic data processing algorithms, for two main reasons: first, the amount of data to be analyzed is tremendous and cannot be exploited efficiently without tools to detect, summarize and present in a suitable way only the most relevant information from the bulk of data; second, the raw video data may not be displayed completely as privacy issues would require the data to be either adequately filtered to select only a subset of the video or processed entirely automatically to produce aggregated descriptors.

In [3] the framework for video monitoring with wearable camera was designed. The recording is realized at patient's home, i.e. in an ecological situation. The computer vision task consists in an automatic recognition and indexing of IADLs from a taxonomy proposed to patients in each recording.

In this paper, we focus on two technology modules within this framework: automatic estimation of localization and visual saliency prediction of the patient .executing the IADL

## II. MATERIAL AND METHODS

*Wearable camera acquisition Hardware*

The camera is fixed on a lightweight jacket worn by the person. It has to be positioned such that the field of view captures the instrumental activity zone in front of the person and context elements without being occluded by the shoulder. The acquisition set-up was studied by neuro-psycho motriciens in order to ensure the least inconvenience for the patients.

The currently used camera is a GoPro camera that records the video sequence using H.264 compression with a resolution of 1280x960p at 30 frames/s. Its very low weight (around 100g) is unnoticeable by the wearer. During capture, the video is stored on the onboard SD-card, which can then be processed and visualized on a computer with the analysis software.

*Location estimation*

For recognition of IADL and feed-back loop to the patients, the identification of patients' position in his home is of primarily importance. Location estimation from ambient cameras in an instrumented home is not sufficiently precise, due to the presence of occlusions by elements of the environments and caregivers. This is why location estimation from egocentric video is needed.

---
[1] http://www.demcare.eu/

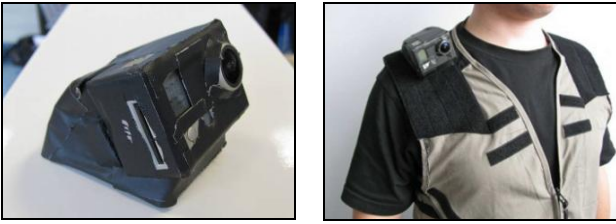

FIGURE 1: Wearable camera device and matching support jacket.

Location estimation can be done at different granularities: room [1], place or 3D position of the patient [2]. When the indoor environment can be modeled by a set of known places or a 3D model the 3D localization of the person can be estimated. The pipeline currently developed and used at University Bordeaux 1 is following the architecture described in [2]. It combines interest point detection and matching with Structure from Motion reconstruction to produce 3D point-clouds from a set of up to hundreds of images.

*Visual saliency for content interpretation*

The content recorded with wearable cameras is a rich source of information for a medical practitioner for analysis of the behavior and the sequence of tasks of the patient executing IADLs. One of the important indicators of the patient's condition is the concentration on the activity being fulfilled. This indicator can be extracted from video, if the focus of interest of the patient differs from predicted one. Furthermore, we seek to limit the automatic analysis of the observed dynamic scene to the area of interest which is visually salient for the medical practitioner. For egocentric video the problem of visual saliency detection is posed in a new and challenging way. Two different persons are involved: the Actor wearing the video camera and the Viewer who is interpreting the video. Their visual saliencies are not the same. Indeed according to the physiological studies [6, 7], the human gaze anticipates the motor action of limbs when fulfilling an activity. We study relation between the Actor's and Viewer's saliencies realized on a subjective saliency maps, obtained with eye-trackers on the basis of this biophysical fundings.

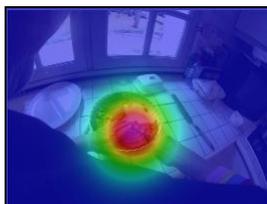

FIGURE 2: Example of visual saliency map focused on the hand of the person wearing the camera.

**Video Data**
For this work, a dataset containing the eye locations of the persons performing the actions (Actors) is needed in order to compare with the gaze coordinates of the people watching these actions on video (Viewers).
Along with their paper [5], the authors have publicly released the GTEA video and gaze dataset obtained with Tobii eye-tracking glasses. The videos are at a 15fps rate and a 640*480 pixel resolution. For the gaze location, two points per frame are recorded (30 samples per second). The subjects were asked to execute a specific IADL "preparing a meal" using different ingredients placed on the table in front of them. In total 17 videos of 4 min in average are available, performed by 14 different participants.

**Extraction of visual saliency maps**
We extract saliency maps from eye-tracker data (subjective maps), both for Actor, supplied with GTEA data set and for the Viewer, which we recorded in a psycho-visual experiment involving 22 subjects. To build the maps, the reference method [10] was used, transforming sparse gaze locations into a dense map by adapted Gaussian filtering. An example of such a map is given in Figure 2. The relation between Actors' and Viewers' saliency maps we propose is based on the findings [6, 7] that there exists a time shift around 500 ms between the gaze fixation on the visual target and beginning of the action. This is the Actor point of view. The Viewer is interested in the action thus his saliency map has to be shifted in time. Hence, our assumption is that given a saliency map of the Viewer, the prediction of the unknown Actor's saliency map is possible by a time shift. In the experiments we verify this hypothesis.

## III. RESULTS

*Location estimation*

The 3D reconstruction approach was tested on data acquired within the Dem@care project in the CHU Nice Laboratory, which is used for assessment on patients. The The current approach is suitable to model selected parts of the environment, which can be reconstructed separately (see Figure 3). This illustrates the possibility to obtain a precise 3D localization of the person in a known environment, to be later merged with complementary information: gaze analysis, location of objects, instrumental actions with the hands…

*Saliency models comparison*

To check the hypothesis about the time-shift between visual saliency (gaze fixation) maps of Viewer and Actor in the egocentric video, we fulfilled the manual test first. The experience was done for 8 of the videos provided from GTEA dataset. The experimental distribution of

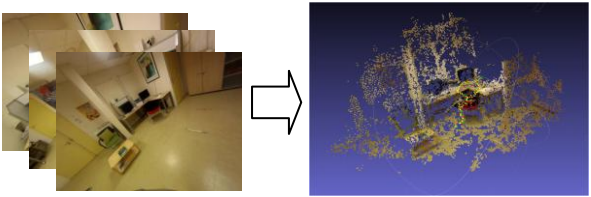

FIGURE 3. Reconstruction of a 3D model of the TV place of the CHUN@Lab environment from GoPro Images. The camera trajectory is shown in green dots (best viewed in color).

Time-shift values showed that most of the actions are acknowledged by the viewer around 8 frames later than the actor (533ms corresponding to [6, 7].)
Then an automatic comparison of visual saliency maps between Viewers and Actors was realized. The similarity metrics used were NSS, AUC, and PCC [9]. They express the correlation between maps. Statistical Wilcoxon test confirmed the hypothesis on the existence of such a time-shift. As for the values, the results with the AUC metric are given by Figure 4. They are displayed for different time-shifts between actors' (fixed) and viewers' (varying in time) saliency maps. The NSS and PCC comparison metrics scores behaved similarly. The computation of these three metrics clearly brings to the same conclusion: the actors' saliency maps show more correspondence with those of the viewers when the latter are considered with an advance (around 10 frames = 667ms). Hence the Actor saliency can be predicted from that one of the viewer.

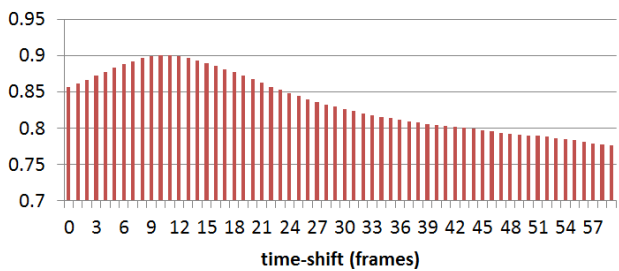

*FIGURE 4*- Histogram of AUC scores between the actor's and viewer's saliency maps for different time-shifts

## IV.  DISCUSSION - CONCLUSION

Hence in this paper we introduced a new an exciting approach in the use of IT for Alzheimer disease assessment and studies: the egocentric video. The results of our anterior studies confirmed that such video observations allow for objective assessment of patients IADL [3]. The proposed methods for automatic analysis, such as location estimation of the patient and prediction of visual saliency, that is of the focus of interest of the patient, allow for identifying important elements in patient's behavior when executing IADL: i) where he is and ii) where he is supposed to look. The proposed methods are being integrated in the whole automatic system of patients' behavior designed in the Dem@care project. Additional recordings are currently planned within the Dem@care project, in both controlled scenarios in Lab, as well as in Homes.

ACKNOWLEDGEMENT

This research has been supported by national grant ANR-09-BLAN-0165-02 (IMMED project) and the European Community's program (FP7/2007–2013) under Grant Agreement 288199 (Dem@Care project).

REFERENCES

[1] V. Dovgalecs, R. Mégret, and Y. Berthoumieu, "Multiple Feature Fusion Based on Co-Training Approach and Time Regularization for Place Classification in Wearable Video," Advances in Multimedia, vol. 2013, Article ID 175064, 22 pages, 2013.

[2] H. Wannous, V. Dovgalecs, R. Mégret, M. Daoudi: Place Recognition via 3D Modelling for Personal Activity Lifelog Using Wearable Camera. Advances in Multimedia Modelling - 18th International Conference, MMM 2012, Klagenfurt, Austria, 2012, pp244-254.

[3] Karaman, S., Benois-Pineau, J., Mégret, R., Dovgalecs, V., Dartigues, J.F., Gäestel, Y.: Human Daily Activities Indexing in Videos from Wearable Cameras for Monitoring of Patients with Dementia Diseases. In: ICPR 2010

[4]. D. Szolgay, J. Benois-Pineau, R. Mégret, Y. Gäestel, J.F. Dartigues: Detection of moving foreground objects in videos with strong camera motion. Pattern Analysis and Applications 14 (2011) 311–328

[5] A. Fathi, Y. Li, J. M. Rehg, "Learning to recognize daily actions using gaze," in ECCV 2012, vol. 7572 of Lecture Notes in Computer Science, pp. 314–327.

[6] C. Prablanc, J.F. Echailler, E. Komilis, and M. Jeannerod, "Optimal response of eye and hand motor systems in pointing at a visual target," Biol. Cybernetics, vol. 35, pp. 113–124, 1979.

[7] M. Dorr, T. Martinetz, K. R. Gegenfurtner, and E. Barth, "Variability of eye movements when viewing dynamic natural scenes.," Journal of vision, vol. 10, no. 10, 2010.

[8] M. F. Land and M. Hayhoe, "In what ways do eye movements contribute to everyday activities?," Vision research, vol. 41, no. 25-26, pp. 3559–3565, 2001.

[9] O. Le Meur and T. Baccino, "Methods for comparing scanpaths and saliency maps: strengths and weaknesses.," Behav Res Methods, 2012.

[10] D. Wooding, "Eye movements of large populations: deriving regions of interest, coverage, and similarity using fixation maps," Behavior Research Methods, vol 34, pp. 518–528, 2002